\documentclass[10pt, conference]{IEEEtran}

\usepackage{amsmath}
\usepackage{amssymb}
\usepackage{verbatim}
\usepackage{epsfig}
\usepackage{subfigure}

\begin{document}

%\title{On Capacity-Optimal Frequency Slot Size for Ad Hoc Networks}
\title{Optimizing the SINR operating point of spatial networks}

\author{
\authorblockN{Nihar Jindal}
\authorblockA{
ECE Department \\
University of Minnesota \\
%Minneapolis, MN 55455 \\
nihar@umn.edu} \and
\authorblockN{Jeffrey G. Andrews}
\authorblockA{
ECE Department \\
University of Texas at Austin \\
% 2540 Dole Street \\
%Austin, TX  \\
jandrews@ece.utexas.edu} \and
\authorblockN{Steven Weber}
\authorblockA{
ECE Department  \\
Drexel University \\
%Philadelphia, PA \\
sweber@ece.drexel.edu } }

% make the title area
\maketitle

\begin{abstract}
This paper addresses the following question, which is of interest in
the design and deployment of a multiuser decentralized network.
Given a total system bandwidth of $W$ Hz and a fixed data rate
constraint of $R$ bps for each transmission, how many frequency
slots $N$ of size $W/N$ should the band be partitioned into to
maximize the number of simultaneous transmissions in the network?
%For a centralized (e.g. cellular) system in isolation, the answer is
%trivially $N=1$, since the received
%signal-to-interference-plus-noise ratio (SINR) must increase
%exponentially with $N$ to maintain the rate $R$, and so dividing the
%band is counter-productive assuming a transmit power constraint.  In
%an ad hoc network, however, dividing up the band allows users access
%to multiple orthogonal bands, which provides the potential for a
%network density increase.  On the other hand, each user must now
%receive a higher SINR, which limits the network density, since
%denser networks result in more interference.
In an interference-limited ad-hoc network, dividing the available
spectrum results in two competing effects: on the positive side, it
reduces the number of users on each band and therefore decreases the
interference level which leads to an increased SINR, while on the
negative side the SINR requirement for each transmission is
increased because the same information rate must be achieved over a
smaller bandwidth.   Exploring this tradeoff between bandwidth and
SINR and determining the optimum value of $N$ in terms of the system
parameters is the focus of the paper. Using stochastic geometry, we
analytically derive the optimal SINR threshold (which directly
corresponds to the optimal spectral efficiency) on this tradeoff
curve and show that it is a function of only the path loss exponent.
Furthermore, the optimal SINR point lies between the low-SINR
(power-limited) and high-SINR (bandwidth-limited) regimes.  In order
to operate at this optimal point, the number of frequency bands
(i.e., the reuse factor) should be increased until the threshold
SINR, which is an increasing function of the reuse factor, is equal
to the optimal value.

\end{abstract}

\section{Introduction}
We consider a spatially distributed network, representing either a
wireless ad hoc network or unlicensed (and uncoordinated) spectrum
usage by many nodes (e.g., WiFi), and consider the tradeoff between
bandwidth and SINR. We ask the following question: given a fixed
total system bandwidth and a fixed rate requirement for each
single-hop transmitter-receiver link in the network, at what point
along the bandwidth-SINR tradeoff-curve should the system operate at
in order to maximize the spatial density of transmissions subject to
an outage constraint?  Note that the outage probability is computed
with respect to random user locations as well as fading.

For example, given a system-wide bandwidth of 1 Hz and a desired
rate of 1 bit/sec, should  (a) each transmitter utilize the entire
spectrum (e.g., transmit one symbol per second) and thus require an
SINR of 1 (utilizing $R=W \log(1 + SINR)$ if interference is treated
as noise), (b) the band be split into two orthogonal 0.5 Hz sub-bands
where each transmitter utilizes one of the sub-bands with the
required SINR equal to 3, or (c) the band be split into $N > 2$
orthogonal $\frac{1}{N}$ Hz sub-bands where each transmitter
utilizes one of the sub-bands with the required SINR equal to
$2^N-1$? Note that an equivalent formulation is to optimize the
bandwidth-SINR operating point such that the outage probability is
minimized for some fixed density of transmissions.

%Note that this is somewhat similar to the problem of optimizing
%frequency reuse in cellular networks (*refs*) applied to spatially
%random networks.

%minimize the outage
%probability, or equivalently to maximize the allowable (spatial)
%intensity of transmissions such that an outage constraint is
%not violated?

We consider a network with the following key characteristics:
\begin{itemize}
    \item Transmitter node locations are a realization of a homogeneous
    spatial Poisson process.
    \item Each transmitter communicates with a single receiver
that is a reference distance $d$ meters away.
    \item All transmissions are constrained to have an absolute
    rate of $R$ bits/sec regardless of the bandwidth.
    \item All multi-user interference is treated as noise.
    \item The channel is frequency-flat, reflects path-loss and
possibly fast and/or slow fading, and is constant for the duration
of a transmission.
    \item Transmitters do not have channel state information and no
transmission scheduling is performed, i.e., transmissions are
independent and random, conceptually like an Aloha system.
\end{itemize}
The last assumption should make it clear that we are considering
only an \textit{off-line} optimization of the frequency band
structure, and that no on-line (e.g., channel- and queue-based)
transmission or sub-band decisions are considered.

 These assumptions are chosen primarily for tractability and
their validity will not be assured in all implementations, but
generalizations are left to future work.

\subsection{Related Work}

The transmission capacity framework introduced in \cite{WebYan05}
is used to quantify the throughput of such a network, since this
metric captures notions of spatial density, data rate, and outage
probability, and is more amenable to analysis than the more
popular transport capacity \cite{GupKum00}. Using tools from
stochastic geometry \cite{StoKen96}, the distribution of
interference from other concurrent transmissions at a reference
receiving node\footnote{The randomness in interference is only due
to the random positions of the interfering nodes and fading.} is
characterized as a function of the spatial density of
transmitters, the path-loss exponent, and possibly the fading
distribution.  The distribution of SINR at the receiving node can
then be computed, and an outage occurs whenever the SINR falls
below some threshold $\beta$. The outage probability is clearly an
increasing function of the density of transmissions, and the
transmission capacity is defined to be the maximum density of
successful transmissions such that the outage probability is no
larger than some prescribed constant $\epsilon$.

The problem studied in this work is essentially the optimization of
frequency reuse in uncoordinated spatial (ad hoc) networks, which is
a well studied problem in the context of cellular networks (see for
example \cite{Rappaport} and references therein). In both settings
the tradeoff is between the bandwidth  utilized per
cell/transmission, which is inversely proportional to the frequency
reuse factor, and the achieved SINR per transmission. A key
difference is that in cellular networks, regular frequency reuse
patterns can be planned and implemented, whereas in an ad hoc
network this is impossible and so the best that can be hoped for is
uncoordinated \textit{random} frequency reuse.  Another crucial
difference is in terms of analytical tractability.  Although there
has been a tremendous amount of work on optimization of frequency
reuse for cellular networks, these efforts do not, to the best of
our knowledge, lend themselves to clean analytical results.  On the
contrary, in this work we are able to derive very simple analytical
results in the random network setting that very cleanly show the
dependence of the optimal reuse factor on system parameters such as
path loss exponent and rate.

Perhaps the most closely related work is
\cite{Sikora_Laneman_Haenggi}\cite{Sikora_Laneman_Haenggi_ITW04},
in which a one-dimensional (i.e., linear), evenly spaced,
multi-hop wireless network is studied.  In finding the optimal (in
terms of total energy minimization) number of intermediate relay
nodes in an interference-free network (i.e., each hop is assigned
a distinct frequency or time slot), their analysis (rather
remarkably) coincides almost exactly with our analysis of an
interference-limited, two-dimensional, random network.  The issue
of frequency reuse in interference-limited 1-D networks is also
explicitly considered in \cite{Sikora_Laneman_Haenggi}, and some
of the general insights are similar to those derived in this work.

%We comment further on these similarities in Section
%\ref{sec-opt_freq}.

\section{Key Insights}

The bandwidth-SINR tradeoff reveals itself if the system bandwidth
is split into $N$ non-overlapping bands and each transmitter
transmits on a randomly chosen band with some fixed power
(independent of $N$).  This splitting of the spectrum results in
two competing effects.  First, the density of transmitters on each
band is a factor of $N$ smaller than the overall density of
transmitters, which reduces interference and thus increases SINR.
Second, the threshold SINR must be increased in order to maintain
a fixed rate while transmitting over $\frac{1}{N}$-th of the
bandwidth.  This requires a reduced network density in order to
meet the prescribed outage constraint.

Although intuition from point-to-point AWGN channels -- for which
capacity is a strictly increasing function of bandwidth if
transmission power is fixed -- might cause one to think that the
optimum solution is trivially to not split the band ($N=1$), this is
generally quite far from the optimum in ad hoc networks. Our
analysis shows that $N$ should be chosen such that the required
threshold SINR lies between the low-SNR (power-limited) and high-SNR
(bandwidth-limited) regimes, for example in the range of 0 - 5 dB
for reasonable path loss exponents.  This approximately corresponds
to the region where the function $\log(1 + SINR)$ transitions from
linear to logarithmic in SINR.

The intuition behind this result is actually quite simple: if $N$ is
such that the threshold SINR is in the wideband regime (roughly
speaking, below 0 dB), then doubling $N$ leads to an approximate
doubling of the threshold SINR, or equivalently a 3 dB increase.
Whenever the path-loss exponent is strictly greater than 2, doubling
the threshold SINR reduces the allowable intensity of transmissions
on each band by a factor strictly smaller than two. On the other
hand, the doubling of $N$ increases the total intensity by exactly a
factor of two because the number sub-bands is increased by the a
factor of two; the combination of these effects is a net increase in
the allowable intensity of transmissions. Therefore, it is
beneficial to continue to increase $N$ until the point at which the
required SINR threshold begins to increase \textit{exponentially}
rather than \textit{linearly} with $N$.

%In a point-to-point AWGN channel, reducing bandwidth results in: (a)
%reduced noise power and thus an increased SNR (assuming fixed
%transmit power), and (b) reduced degrees of freedom (e.g., a reduced
%symbol rate).  Indeed, the increasing nature of capacity with
%bandwidth indicates that this tradeoff is never a worthwhile one:
%the cost of the reduced degrees of freedom always outweighs the
%benefit of the increased SNR.

%In fact, our approach is not even applicable in this case since in a
%point-to-point channel, it is not possible to divide the band to
%create $N$ channels each with the same data rate as the original
%channel unless the transmit power is increased by a factor of
%$O(2^N)$, which is expressly prohibited in our model.  In an
%interference-limited network though, the SINR can be increased
%without an increase in transmit power by reducing the amount of
%interference, at the cost of bandwidth.

%In an interference-limited multi-user network, decreasing the
%bandwidth utilized by each transmission also results in increased
%SINR (primarily due to the reduction in interference rather than
%noise) and reduced degrees of freedom (which has the effect of
%increasing the SINR threshold in our fixed rate setting).  However,
%if one is in in the wideband (or power-limited) regime, this is
%actually a worthwhile tradeoff to make.

\section{Preliminaries}

\subsection{System Model}
We consider a set of transmitting nodes at an arbitrary snapshot
in time with locations specified by a homogeneous Poisson process
of intensity $\lambda$ on the infinite two-dimensional plane.  We
consider a reference receiver that is located, without loss of
generality, at the origin, and let $X_i$ denote the distance of
the $i$-th transmitting node to the reference receiver.  The
reference transmitter is placed a fixed distance $d$ away.
Received power is modelled by path loss with exponent $\alpha > 2$
and a distance-independent fading coefficient $h_i$ (from the
$i$-th transmitter to the reference receiver).  Therefore, the
SINR at the reference receiver is:
\begin{eqnarray*}
SINR_0 = \frac{ \rho d^{-\alpha} |h_0| }{ \eta + \sum_{i \in
\Pi(\lambda)} \rho X_i^{-\alpha} |h_i| },
\end{eqnarray*}
where $\Pi(\lambda)$ indicates the point process describing the
(random) interferer locations, and $\eta$ is the noise power.  If
Gaussian signalling is used by all nodes, the mutual information
conditioned on the transmitter locations and the fading realizations
is:
\begin{eqnarray*}
I(X_0;Y_0 | \Pi(\lambda), \mathbf{h}) = \log_2 ( 1 + SINR_0),
\end{eqnarray*}
where $\mathbf{h} = (h_0,h_1,\ldots)$.  Notice that we assume that
all nodes simultaneously transmit with the same power $\rho$,
i.e., power control is not used. Moreover, nodes decide to
transmit independently and irrespective of their channel
conditions, which corresponds roughly to slotted ALOHA (i.e.,  no
scheduling is performed).

A few comments in justification of the above model are in order.
Although the model contains many simplifications to allow for
tractability, it contains many of the critical elements of a real ad
hoc network.  First, the spatial Poisson distribution means that
nodes are randomly and independently located; this is reasonable
particularly in a network with substantial mobility or
indiscriminate node placement (for example a very dense sensor
network).  The fixed transmission distance of $d$ is clearly not a
reasonable assumption; however our prior work
\cite{WebYan05,WebAndJin06} has shown rigorously that variable
transmit distances do not result in fundamentally different capacity
results, so a fixed distance is chosen because it is much simpler
analytically and allows for crisper insights.  A similar
justification can be given for ignoring power control, although
power control is often not used in actual ad hoc networks either.
Finally, scheduling procedures (e.g., using carrier sensing to
intelligently select a sub-band) may significantly affect the
results and is definitely of interest, but this opens many more
questions and so is left to future work.

\subsection{Transmission Capacity Model}

In the outage-based transmission capacity framework, an outage
occurs whenever the SINR falls below a prescribed threshold
$\beta$, or equivalently whenever the instantaneous mutual
information falls below $\log_2(1+ \beta)$.  Therefore, the
system-wide outage probability is:
\begin{eqnarray*}
P \left( \frac{ \rho d^{-\alpha} |h_0| }{ \eta + \sum_{i \in
\Pi(\lambda)} \rho X_i^{-\alpha} |h_i| } \leq \beta \right).
\end{eqnarray*}
This quantity is computed over the distribution of transmitter
positions as well as the iid fading coefficients, and thus
corresponds to fading that occurs on a slower time-scale than packet
transmission. The outage probability is clearly an increasing
function of the intensity $\lambda$.

If $\lambda(\epsilon)$ is the maximum intensity of
\textit{attempted} transmissions such that the outage probability
(for a fixed $\beta$) is no larger than $\epsilon$, then the
transmission capacity is then defined as $c(\epsilon) =
\lambda(\epsilon) (1 - \epsilon) b$, which is the maximum density
of \textit{successful} transmissions times the spectral efficiency
$b$ of each transmission.  In other words, transmission capacity
is like area spectral efficiency subject to an outage constraint.
Using tools from stochastic geometry, in \cite{WebYan05} it is
shown that the maximum spatial intensity $\lambda(\epsilon)$ for
small values of $\epsilon$ is:
\begin{eqnarray} \label{eq-transcap}
\lambda(\epsilon) = \frac{c}{\pi d^2} \left( \frac{1}{\beta} -
\frac{\eta}{\rho d^{-\alpha}} \right)^{\frac{2}{\alpha}} \epsilon +
O(\epsilon^2),
\end{eqnarray}
where $c$ is a constant that depends only on the distribution of the
fading coefficients \cite{WebAndJin06}.
%where $c$ is a constant satisfying $1 - \frac{1}{\alpha} \leq c \leq 1$.
In the proceeding analysis, the key is the manner in which the
transmission capacity varies with the SINR constraint $\beta$; for
small noise values, which is the case in the interference-limited
scenarios we are most interested in, intensity is proportional to
$\beta^{-\frac{2}{\alpha}}$.  Because fading has only a
multiplicative effect on transmission capacity, it does not effect
the SINR-bandwidth tradeoff and thus is not considered in the
remainder of the paper.

\section{Optimizing Frequency Usage} \label{sec-opt_freq}

In this section we consider a network with a fixed total bandwidth
of $W$ Hz, and where each link has a rate requirement of $R$
bits/sec and an outage constraint $\epsilon$. Assuming the network
operates as described in the previous section, the goal is to
determine the optimum number of sub-bands $N$ into which the
system bandwidth of $W$ Hz should be divided while meeting these
criteria.  By optimum, we mean the choice of $N$ that maximizes
the intensity of allowable transmissions $\lambda(\epsilon,N)$. As
we will see, due to our constraint that the data rate on each link
is the same regardless of $\lambda$ and $N$, this also corresponds
to maximizing transmission capacity.

\subsection{Definitions and Setup}

In performing this analysis, we assume that there exist coding
schemes that operate at any point along the AWGN capacity
curve.\footnote{In Section \ref {sec-capacity_gap} we relax this
assumption by allowing for operation at a constant coding gap
(i.e., power offset) from AWGN capacity, and see that this has no
effect on our analysis.} To facilitate exposition, we define the
\textit{spectral utilization} $\tilde{R}$ to be the ratio of the
required rate relative to the total system bandwidth:
\begin{eqnarray*}
\tilde{R} \triangleq \frac{R}{W} ~~ \textrm{bps/Hz/user}.
\end{eqnarray*}
Note that we intentionally refer to $\tilde{R}$, which is
externally defined, as the spectral utilization; the
\textit{spectral efficiency}, on the other hand, is a system
design parameter determined by the choice of $N$.

If the system bandwidth is not split ($N=1$), each node utilizes the
entire bandwidth of $W$ Hz. Therefore, the required SINR $\beta$ is
determined by inverting the standard rate expression:
\begin{eqnarray*}
R = W \log_2 (1 + \beta),
\end{eqnarray*}
which gives $\beta = 2^{\frac{R}{W}} - 1 = 2^{\tilde{R}} - 1$. The
maximum intensity of transmissions can be determined by plugging
in this value of $\beta$ into (\ref{eq-transcap}), along with the
other relevant constants.

More generally, if the system bandwidth is split into $N$
orthogonal sub-bands each of width $\frac{W}{N}$, and each
transmitter-receiver pair uses only one of these sub-bands at
random, the required SINR $\beta(N)$ is determined by inverting
the rate expression:
\begin{eqnarray*}
R &=& \frac{W}{N} \log_2 (1 + \beta(N)),
\end{eqnarray*}
which yields:
\begin{eqnarray*}
\beta(N) &=& 2^{\frac{N R}{W}} - 1 = 2^{N\tilde{R}} - 1.
\end{eqnarray*}
Notice that the spectral efficiency (on each sub-band) is $b =
\frac{R}{W/N}$ bps/Hz, which is $N$ times the spectral utilization
$\tilde{R}$. The maximum intensity of transmissions \textit{per
sub-band} for a particular value of $N$ is determined by plugging
$\beta(N)$ into (\ref{eq-transcap}) with noise power $\eta =
\frac{W}{N} N_0$. Since the $N$ sub-bands are statistically
identical, the maximum total intensity of transmissions, denoted by
$\lambda(\epsilon,N)$, is the per sub-band intensity multiplied by a
factor of $N$.  Dropping the second order term in
(\ref{eq-transcap}), we have:
\begin{eqnarray} \label{eq-tc_approx1}
\lambda(\epsilon,N) \approx N \left( \frac{\epsilon}{\pi d^2}
\right) \left( \frac{1}{\beta(N)} - \frac{1}{N\cdot SNR}
\right)^{\frac{2}{\alpha}},
\end{eqnarray}
where the constant  $SNR \triangleq \frac{\rho d^{-\alpha}}{N_0 W}$
is the signal-to-noise ratio in the absence of interference when the
entire band is used.

\subsection{Optimization}

Optimizing the number of sub-bands $N$ therefore reduces to the
following one-dimensional maximization:
\begin{eqnarray} \label{eq-optN}
N^* = \textrm{arg} \max_N \lambda(\epsilon,N),
\end{eqnarray}
which yields a solution that depends only on the path-loss exponent
$\alpha$, the spectral utilization $\tilde{R}$, and the constant
$SNR$.

In general, the interference-free $SNR$ can be ignored because the
systems of interest are interference- rather than noise-limited.
Assuming $SNR$ is infinite we have:
\begin{eqnarray}
\lambda(\epsilon, N) &\approx&  \left( \frac{\epsilon}{\pi d^2}
\right) N \cdot \beta(N)^{-\frac{2}{\alpha}} \\
&=&  \left( \frac{\epsilon}{\pi d^2} \right) N (2^{N\tilde{R}}
-1)^{-\frac{2}{\alpha}}. \label{eq-tc_approx_fh}
%&=&  \left( \frac{\epsilon}{\pi d^2} \right) N (e^{N \tilde{R}_e}
%-1)^{-\frac{2}{\alpha}}, \label{eq-tc_ln}
\end{eqnarray}
%where we have changed to natural logarithm for convenience and
% $\tilde{R}_e = \tilde{R} \log_2 e$ is the spectral utilization
%in nats/Hz rather than bps/Hz.
The leading factor of $N$ represents the fact that total
transmission intensity is $N$ times the per-band intensity, while
the $(2^{N \tilde{R}} -1)^{-\frac{2}{\alpha}}$ term, which is a
decreasing function of $N$, is the amount by which intensity must be
decreased in order to maintain an outage probability no larger than
$\epsilon$ in light of the monotonically increasing (in $N$) SINR
threshold $\beta(N)$.

Since $\tilde{R}$ is a constant, we can make the substitution $b=
N \tilde{R}$ and equivalently maximize the function
$b(2^b-1)^{-\frac{2}{\alpha}}$.  Taking the derivative with
respect to $b$ we get:
\begin{eqnarray*}
\lefteqn{\frac{\partial}{\partial b} \left[
b(2^b-1)^{-\frac{2}{\alpha}} \right]
%&=& (2^b-1)^{-\frac{2}{\alpha}} + b
%\left(-\frac{2}{\alpha}\right)  2^b (2^b-1)^{-\frac{2}{\alpha}-1} \log_e 2 \\
%&=& (2^b-1)^{-\frac{2}{\alpha}}
%\left[1 - \frac{2}{\alpha}b e^b (2^b-1)^{-1} \log_e 2 \right] \\
=} \\ && (2^b-1)^{-\frac{2}{\alpha}} \left[1 - \frac{2}{\alpha}b
(1-2^{-b})^{-1} \log_e 2 \right].
\end{eqnarray*}
Since the first term is strictly positive for $b > 0$, we set the
second term to zero to get a fixed point equation for the optimal
spectral efficiency $b^*$:
\begin{eqnarray} \label{eq-optspec}
%1 - \frac{2}{\alpha}b (1-e^{-b})^{-1} = 0 ~~ \rightarrow ~~
b^* = (\log_2 e) \frac{\alpha}{2} (1 - e^{-b^*}),
\end{eqnarray}
which has solution
\begin{equation} \label{eq-optspec2}
b^* = \log_2 e \left[ \frac{\alpha}{2} + W \left( -
\frac{\alpha}{2} e^{-\frac{\alpha}{2}} \right) \right],
\end{equation}
where $W(z)$ is the principle branch of the Lambert $W$ function
and thus solves $W(z) e^{W(z)} = z$.\footnote{Equation
(\ref{eq-optspec2}) is nearly identical, save for a factor of 2,
to the expression for the optimal number of hops in an
interference-free linear network given in equation (18) of
\cite{Sikora_Laneman_Haenggi}.  This similarity is due to the fact
that the objective function in equation (17) of
\cite{Sikora_Laneman_Haenggi} coincides almost exactly with
(\ref{eq-tc_approx_fh}).}

Because $1 - \frac{2}{\alpha}b (1-2^{-b})^{-1} \log_e 2$ is strictly
decreasing and $(2^b-1)^{-\frac{2}{\alpha}}$ is strictly positive,
the first derivative is strictly positive for $0 < b < b^*$ and is
strictly negative for $b > b^*$. Therefore, the $b^*$ in
(\ref{eq-optspec2}) is indeed the unique maximizer.  Furthermore, it
is easily shown that the optimizing $b^*$ is an increasing function
of $\alpha$, is upper bounded by $\frac{\alpha}{2} \log_2 e$, and
that $b^*/(\frac{\alpha}{2} \log_2 e)$ converges to $1$ as $\alpha$
grows large.

Recalling that $b = N \tilde{R}$ is the spectral efficiency on
each sub-band, the quantity $b^*$, which is a function of only the
path-loss exponent $\alpha$, is the \textit{optimum spectral
efficiency}.\footnote{An optimal spectral efficiency is derived
for interference-free, regularly spaced, 1-D networks in
\cite{Sikora_Laneman_Haenggi_ITW04}; however, these results differ
by approximately a factor of 2 from our results due to the
difference in the network dimensionality.}  Therefore, the optimal
value of $N$ (ignoring the integer constraint) is determined by
simply dividing the optimal spectrum efficiency $b^*$ by the
spectral utilization $\tilde{R}$:
\begin{equation} \label{eq-Nstar}
N^* = \frac{b^*}{\tilde{R}}.
\end{equation}
To take care of the integer constraint on $N$, the nature of the
derivative of $b(2^b-1)^{-\frac{2}{\alpha}}$ makes it sufficient to
consider only the integer floor and ceiling of $N^*$ in
(\ref{eq-Nstar}).  Note that the optimal number of sub-bands depends
only on the spectral utilization $\tilde{R}$ (inversely) and on
$b^*$, which is a function of the path-loss exponent; there is no
dependence on either the outage constraint $\epsilon$ or on the
transmission range $d$.

If the spectral utilization is larger than the optimum spectral
efficiency, i.e., $\tilde{R} \geq b^*$, then choosing $N=1$ is
optimal.  On the other hand, if $\tilde{R} \leq \frac{1}{2} b^*$,
then the optimal $N$ is strictly larger than 1.  In the
intermediate regime where $\frac{1}{2} b^* \leq \tilde{R} \leq b^*
$, the optimal $N$ is either one or two.

\begin{figure}
\centering
\includegraphics[width=3in]{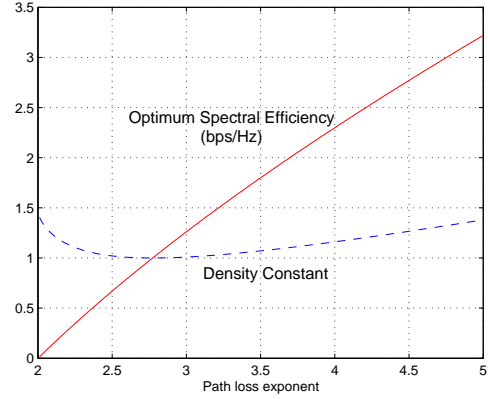}
\caption{Optimal Spectral Efficiency vs. Path-Loss Exponent}
\label{fig-opt_spec}
\end{figure}

In Fig. \ref{fig-opt_spec} the optimal spectral efficiency $b^*$
is plotted (in units of bps/Hz) as a function of the path-loss
exponent $\alpha$, along with the quantity $b^*(2^{b^*}-1)^{-
\frac{2}{\alpha}}$, which is referred to as the density constant
because the optimal density $\lambda^*(\epsilon)$ is this quantity
multiplied by $\left( \frac{\epsilon}{ \tilde{R} \pi d^2}
\right)$. The optimal spectral efficiency is very small for
$\alpha$ close to 2 but then increases nearly linearly with
$\alpha$; for example, the optimal spectral efficiency for
$\alpha=3$ is $1.26$ bps/Hz (corresponding to $\beta=1.45$ dB).

%\begin{figure}
%\centering
%\epsfig{file=opt_spec.eps, width=3in}
%\caption{Optimal Spectral Efficiency vs. Path-Loss Exponent}
%\label{fig-opt_spec}
%\end{figure}

\subsection{Interpretation}

To gain an intuitive understanding of the optimal solution, first
consider the behavior of $\lambda(\epsilon, N)$ when the quantity
$N \tilde{R}$ is small, i.e. $N \tilde{R} \ll 1$.  In this regime,
the SINR threshold $\beta(N)$ grows approximately linearly with
$N$:
\begin{eqnarray*}
\beta(N) = 2^{N \tilde{R}} -1 &=& e^{N \tilde{R} \log_e 2} -1 \\ &=&
\sum_{k=1}^{\infty} \frac{(N \tilde{R} \log_e 2)^k}{k!} \\ &\approx&
N \tilde{R} \log_e 2.
\end{eqnarray*}
Plugging into (\ref{eq-tc_approx_fh}) we have
\begin{eqnarray*}
\lambda(\epsilon, N) &\approx&  \left( \frac{\epsilon}{\pi d^2} \right) N (N \tilde{R} \log_e 2)^{-\frac{2}{\alpha}}  \\
&=& \left( \frac{\epsilon}{\pi d^2} \right) {\tilde{R} \log_e
2}^{-\frac{2}{\alpha}} N^{\left(1-\frac{2}{\alpha} \right)}.
\end{eqnarray*}
For any path-loss exponent $\alpha > 2$, the maximum intensity of
transmissions monotonically increases with the number of sub-bands
$N$ as $N^{\left(1-\frac{2}{\alpha} \right)}$, i.e., \textit{using
more sub-bands with higher spectral efficiency leads to an
increased transmission capacity}, as long as the linear
approximation to $\beta(N)$ remains valid.  The key reason for
this behavior is the fact that transmission capacity scales with
the SINR threshold as $\beta^{-\frac{2}{\alpha}}$, which
translates to $N^{-\frac{2}{\alpha}}$ in the low spectral
efficiency regime.

As $N\tilde{R}$ increases, the linear approximation to $\beta(N)$
becomes increasingly inaccurate because $\beta(N)$ begins to grow
\textit{exponentially} rather than linearly with $N$.  In this
regime, the SINR cost of increasing spectral efficiency is
extremely large. For example, doubling spectral efficiency
requires doubling the SINR \textit{in dB units} rather than in
linear units. Clearly, the benefit of further increasing the
number of sub-bands is strongly outweighed by the SINR cost.

Thus, when $N$ is such that the spectral efficiency $N \tilde{R}$
is relatively small (i.e., less than one), $N$ should be increased
because the benefit of reduced interference outweighs the cost of
the increasing SINR threshold. However, as $N \tilde{R}$
increases, the cost of the (exponentially) increasing the SINR
threshold eventually outweighs the benefit of reduced
interference.  Since transmission capacity depends on the SINR
threshold raised to the power $-\frac{2}{\alpha}$, a larger path
loss exponent corresponds to weaker dependence on the SINR
threshold and thus a larger optimum spectral efficiency $b^*$.

\section{Numerical Results and Discussion}

%\begin{figure*}
%\centering
%\mbox{\subfigure[High spectral efficiency]{\epsfig{file=fixed_bw_a.eps, width=80mm}}\qquad\qquad
%\subfigure[Low spectral efficiency]{\epsfig{file=fixed_bw_b.eps, width=80mm}}}
%\caption{Transmission capacity as a function of \# of sub-bands for
%different desired reference spectral efficiencies}
%\label{fig-fixed_bw}
%\end{figure*}

%\begin{figure*}
%\centering
%\mbox{\subfigure[High spectral efficiency]{
%\includegraphics[width=80mm]{fixed_bw_a}}\qquad\qquad
%\subfigure[Low spectral efficiency]{\includegraphics[width=80mm]{fixed_bw_b}}}
%\caption{Transmission capacity as a function of \# of sub-bands for
%different desired reference spectral efficiencies}
%\label{fig-fixed_bw}
%\end{figure*}

\begin{figure}
\centering
\includegraphics[width=3in]{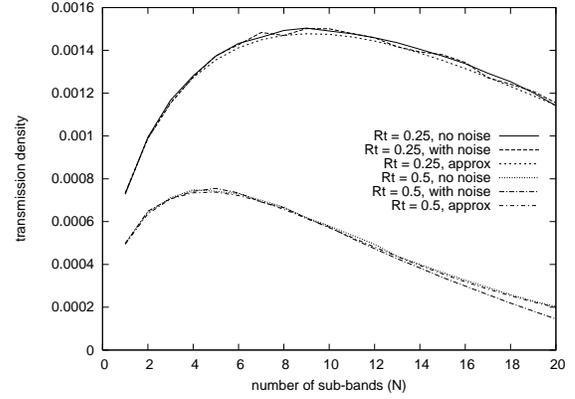}
\caption{Optimal Spectral Efficiency vs. Path-Loss Exponent}
\label{fig-numerical}
\end{figure}

In Figure \ref{fig-numerical}, the maximum density of transmissions
is plotted as a function of $N$ for two different spectrum
utilizations $\tilde{R}$ for a network with $\alpha=4$, $d=10$ m,
and an outage constraint of $\epsilon = 0.1$. The bottom set of
curves correspond to a relatively high utilization of
$\tilde{R}=0.5$ bps/Hz, while the top set corresponds to
$\tilde{R}=0.25$ bps/Hz. Each set of three curves correspond to the
approximation from (\ref{eq-tc_approx1}): $\lambda(\epsilon,N)
\approx N \left( \frac{\epsilon}{\pi d^2} \right)
\beta(N)^{-\frac{2}{\alpha}}$, numerically computed values of
$\lambda(\epsilon,N)$ for $SNR=\infty$, and numerically computed
values for $SNR=20$ dB. For both sets of curves, notice that the
approximation, based on which the optimal value of $N$ was derived,
matches almost exactly with the numerically computed values.
Furthermore, introducing noise into the network has a minimal effect
on the density of transmissions.

For a path loss exponent of $4$, evaluation of (\ref{eq-optspec2})
yields an optimal spectral efficiency of $2.3$ bps/Hz.  When
$\tilde{R}=0.25$, this corresponds to $N^* = \frac{2.3}{0.25}=9.2$
and $N=9$ is seen to be the maximizing integer value.  When
$\tilde{R}=0.5$, we have $N^* = 4.6$ and $N=5$ is the optimal
integer choice.  Note that there is a significant penalty to naively
choosing $N=1$: for $\tilde{R}=0.25$ this leads to a factor of $2$
decrease in density, while for $\tilde{R}=0.5$ this leads to loss of
a factor of $1.5$.

\subsection{Direct Sequence Spread-Spectrum}
Another method of utilizing the bandwidth is to use
direct-sequence spread spectrum with a spreading gain of $N$ and
an information bandwidth of $\frac{1}{N}$ (i.e., a symbol rate of
$\frac{1}{N}$). However, the results of \cite{WebYan05} show that
direct-sequence is (rather significantly) inferior to splitting
the frequency band (FDMA) because it is preferable to avoid
interference (FDMA) rather than to suppress it (DS).  More
concretely, if DS-CDMA is used with completely separate
despreading (assuming a spreading gain of $\frac{1}{N}$) and
decoding, the maximum density of transmissions is approximately
equal to:
\begin{eqnarray}
\lambda(\epsilon,N)^{DS} &\approx&  \left( \frac{\epsilon}{\pi
d^2} \right) N^{\frac{2}{\alpha}} \beta(N)^{-\frac{2}{\alpha}}
\end{eqnarray}
Comparing this with the analogous expression in
(\ref{eq-tc_approx_fh}), we see that the key difference is that DS
results in a leading term of $N^{\frac{2}{\alpha}}$ rather than $N$.
As a result, the density remains constant as $N$ is increased in the
regime where $\beta(N)$ is approximately linear, and decreases with
$N$ once $\beta(N)$ begins to behave exponentially. As a result,
using DS can lead to a considerable performance loss relative to
spreading via frequency orthogonalization.

Another way to understand the inferiority of direct-sequence is the
following: using direct-sequence with a spreading gain of
$\frac{1}{N}$ reduces interference power by a factor of $N$ and
thereby increases the SINR roughly by a factor of $N$.  In the
wideband regime, the SINR threshold $\beta(N)$ increases
approximately linearly with $N$ and thus completely offsets the
value of spreading; as a result the maximum density does not depend
on $N$ in this regime. Beyond the wideband regime, the SINR
threshold $\beta(N)$ increases exponentially with $N$, which clearly
outweighs the linearly increasing SINR provided by the spreading
gain; the maximum density decreases with $N$ in this regime.

\subsection{Below-Capacity Transmission} \label{sec-capacity_gap}
In practical systems, it is not generally possible to signal
precisely at capacity.  One very useful approximation is the
capacity gap metric, where $R = \log_2 (1 + \Gamma \cdot SINR)$ and
$\Gamma \leq 1$ is the (power) gap between the signaling rate and
Shannon capacity.  It is straightforward to see that the gap only
increases the SINR threshold by a multiplicative constant: $\beta(N)
= \frac{1}{\Gamma} \left(2^{N\tilde{R}} - 1 \right)$.  As a result,
the earlier analysis remains unchanged and the optimum spectral
efficiency as well as the optimum number of sub-bands are
independent of the gap $\Gamma$. Indeed, the effect of the coding
gap is simply to reduce the density of transmissions by a factor
$\Gamma^{-\frac{2}{\alpha}}$.

\subsection{Fixed vs. Random Networks}
Our analysis holds for networks in which nodes are \textit{randomly}
located according to a homogeneous 2-D Poisson process.   It would
be interesting to know how this compares with the transmission
capacity for any arbitrary, deterministic, placement of nodes (with
zero outage). By comparing the two, we can determine the penalty
that is paid for by having randomly rather than regularly placed
nodes.

To allow for a fair comparison, we develop bounds on a network in
which $\tilde{R}=b^*$ and thus $N=1$ is optimal. A simple upper
bound on the transmission density can be developed by considering
only the interference contribution of the nearest interferer. The
received SIR, again ignoring thermal noise, is upper bounded by
considering the contribution of only the nearest interferer,
assumed to be a distance $s$ away. The SIR upper bound is thus
given by $\frac{\rho d^{-\alpha}}{\rho s^{-\alpha}} = \left(
\frac{s}{d} \right)^{\alpha}$. The upper bound must be above the
threshold $\beta$ if the actual SINR is above $\beta$, and thus
the following is a necessary condition:
\begin{eqnarray*}
\left(\frac{s}{d} \right)^{\alpha} \geq \beta  ~~ \rightarrow ~~ s
\geq d \beta^{\frac{1}{\alpha}}.
\end{eqnarray*}
Therefore, a necessary but not sufficient condition for meeting
the SINR threshold is that there is no interferer within $d
\beta^{\frac{1}{\alpha}}$ meters of a receiver.  As a result, it
is necessary that an area of $\pi d^2 \beta^{\frac{2}{\alpha}}$
meters$^2$ around each receiver be clear of interferers, which
translates into a density upper bound of $\frac{1}{\pi d^2}
\beta^{-\frac{2}{\alpha}}$.  Since $\beta=2^{b^*}-1$, this gives
\begin{equation}
\lambda^{\mathrm{det}} \leq \frac{1}{\pi d^2}
(2^{b^*}-1)^{-\frac{2}{\alpha}}.
\end{equation}
A lower bound to the optimal density is derived by actually
designing a (infinite) placement of transmitters and receivers.
Indeed, by placing transmitters according to a standard square
lattice and placing receivers on a horizontally shifted version of
this lattice, one can achieve a density within about a factor of two
of the upper bound.

The optimal density bounds should be compared to the density of a
random network with $\tilde{R}=b^*$ found from
(\ref{eq-tc_approx_fh}):
%given by (\ref{eq-inf_dens1}):
\begin{equation}
\lambda^{\mathrm{ran}} \approx \frac{\epsilon}{\pi d^2}
(2^{b^*}-1)^{-\frac{2}{\alpha}}.
\end{equation}
Note that the random density is a factor $\epsilon$ smaller than the
upper bound to the deterministic density.  Thus, when $\epsilon$ is
small, e.g., $\epsilon = 0.1$, there is a rather large penalty
associated with random placement of nodes. This indicates that there
potentially is a very significant benefit to performing localized
transmission scheduling in random networks, assuming that the
associated overhead is not too costly.

%%%%%%%%%%%%%%%%%%%%%%%%%%%%%%%%%%%%%%%%%%%%%%%%%%%%%%%%%%%%%%%%%%%%%%%%%%
\section{Information Density}
An interesting \textit{information density} interpretation can be
arrived at by plugging in the appropriate expressions for the
maximum density of transmissions when the number of sub-bands is
optimized.  By plugging in the optimal value of $N$ (and ignoring
the integer constraint on $N$, which is reasonable when
$\tilde{R}$ is considerably smaller than one) we have:
\begin{eqnarray}
\lambda^*(\epsilon) &=& \max_N ~ \lambda(\epsilon,N) \\
% &\approx& \left(\frac{\epsilon}{\pi d^2} \right) \max_N N
% \beta(N)^{-\frac{2}{\alpha}} \\
 &\approx& \left(\frac{\epsilon}{\pi d^2} \right)
 \frac{1}{\tilde{R}} b^*(2^{b^*}-1)^{-\frac{2}{\alpha}}
 \label{eq-inf_dens1}
\end{eqnarray}
where $b^*$ is defined in (\ref{eq-optspec2}) and the quantity
$b^*(2^{b^*}-1)^{-\frac{2}{\alpha}}$ is denoted as the density
constant in Fig. \ref{fig-opt_spec}. The quantity
$\lambda^*(\epsilon)$ is the maximum allowable spatial density of
attempted transmissions per $m^2$ assuming each transmission occurs
over a distance of $d$ meters at spectral utilization $\tilde{R}$
(i.e., with rate equal to $W \tilde{R}$) and that an outage
constraint of $\epsilon$ must be maintained.

From this expression we can make a number of observations regarding
the tradeoffs between the various parameters of interest.  First
note that density is directly proportional to outage $\epsilon$ and
to the inverse of the square of the range $d^{-2}$.  Thus, doubling
the outage constraint leads to a doubling of density, or inversely
tightening the outage constraint by a factor of two leads to a
factor of two reduction in density.  The quadratic nature of the
range dependence implies that doubling transmission distance leads
to a factor of four reduction in density; this is not surprising
given that the area of the circle centered at the receiver with
radius $d$ is $\pi d^2$.  Perhaps one of the most interesting
tradeoffs is between density and rate: since the two quantities are
 inversely proportional, doubling the rate leads to halving
the density, and vice versa.  Note that this relationship is
directly attributable to the fact that $N^*$ is inversely
proportional to $\tilde{R}$: doubling rate leads to reducing $N^*$
by a factor of two, which reduces total density (across all
sub-bands) by a factor of two.

If we consider the product of density and spectral utilization, we
get a quantity that has units bps/Hz/m$^2$:
\begin{eqnarray}
\lambda^*(\epsilon) \tilde{R}  \approx \left(\frac{\epsilon}{\pi
d^2} \right) b^*(2^{b^*}-1)^{-\frac{2}{\alpha}}
 \label{eq-inf_dens2}
\end{eqnarray}
This quantity is very similar to the \textit{area spectral
efficiency} (ASE) defined in \cite{Alouini_Goldsmith_ASE}. In our
random network setting, the ASE is inversely proportional to the
square of the transmission distance, which is somewhat analogous to
cell radius in a cellular network, and is directly proportional to
the outage constraint.  Since the quantity
$b^*(2^{b^*}-1)^{-\frac{2}{\alpha}}$ does not vary too significantly
with the path-loss exponent (see Fig. \ref{fig-opt_spec}) for
$\alpha$ between 2 and 5, we see that ASE and path-loss exponent are
not very strongly dependent.  Perhaps most interesting is the fact
that the ASE does not depend on the desired rate (assuming $N$ is
optimized for rate).  A random network can support a low density of
high rate transmissions or a high density of low rate transmissions,
or any intermediate point between these extremes.

\section{Conclusion}

In this work we studied bandwidth-SINR tradeoffs in ad-hoc networks
and derived the optimal operating spectral efficiency, assuming that
multi-user interference is treated as noise and that no transmission
scheduling is performed.  A network can operate at this optimal
point by dividing the total available bandwidth into sub-bands sized
such that each transmission occurs on one of the sub-bands at
precisely the optimal spectral efficiency.  As a result, the optimal
number of sub-bands is simply the optimal spectral efficiency (which
is a deterministic function of the path loss exponent) divided by
the normalized (by total bandwidth) rate.

The key takeaway of this work is that an interference-limited ad-hoc
network should operate in neither the wideband (power-limited) nor
high-SNR (bandwidth-limited) regimes, but rather at a point between
the two extremes because this is where the optimal balance between
multi-user interference and bandwidth is achieved. Although we
considered a rather simple network model, we believe that many of
the insights developed here will also apply to more complicated
scenarios, e.g., wideband fading channels and networks in which some
degree of local transmission scheduling is performed.

\bibliographystyle{IEEEtran}
\bibliography{capacity,Andrews}

\begin{thebibliography}{1}
\providecommand{\url}[1]{#1}
\csname url@rmstyle\endcsname
\providecommand{\newblock}{\relax}
\providecommand{\bibinfo}[2]{#2}
\providecommand\BIBentrySTDinterwordspacing{\spaceskip=0pt\relax}
\providecommand\BIBentryALTinterwordstretchfactor{4}
\providecommand\BIBentryALTinterwordspacing{\spaceskip=\fontdimen2\font plus
\BIBentryALTinterwordstretchfactor\fontdimen3\font minus
  \fontdimen4\font\relax}
\providecommand\BIBforeignlanguage[2]{{%
\expandafter\ifx\csname l@#1\endcsname\relax
\typeout{** WARNING: IEEEtran.bst: No hyphenation pattern has been}%
\typeout{** loaded for the language `#1'. Using the pattern for}%
\typeout{** the default language instead.}%
\else
\language=\csname l@#1\endcsname
\fi
#2}}

\bibitem{WebYan05}
S.~Weber, X.~Yang, J.~G. Andrews, and G.~de~Veciana, ``Transmission capacity of
  wireless ad hoc networks with outage constraints,'' \emph{IEEE Trans. on
  Info. Theory}, vol.~51, no.~12, pp. 4091--4102, Dec. 2005.

\bibitem{GupKum00}
P.~Gupta and P.~Kumar, ``The capacity of wireless networks,'' \emph{IEEE Trans.
  on Info. Theory}, vol.~46, no.~2, pp. 388--404, Mar. 2000.

\bibitem{StoKen96}
D.~Stoyan, W.~Kendall, and J.~Mecke, \emph{Stochastic Geometry and Its
  Applications, 2nd Edition}.\hskip 1em plus 0.5em minus 0.4em\relax John Wiley
  and Sons, 1996.

\bibitem{Rappaport}
T.~Rappaport, \emph{Wireless Communications: Principles \& Practice}.\hskip 1em
  plus 0.5em minus 0.4em\relax Prentice Hall, 1996.

\bibitem{Sikora_Laneman_Haenggi}
M.~Sikora, J.~N. Laneman, M.~Haenggi, D.~J. Costello, and T.~Fuja, ``Bandwidth-
  and power-efficient routing in linear wireless networks,'' \emph{IEEE Trans.
  Inform. Theory}, vol.~52, pp. 2624--2633, June 2006.

\bibitem{Sikora_Laneman_Haenggi_ITW04}
------, ``On the optimum number of hops in linear ad hoc networks,'' in
  \emph{Proceedings of IEEE Information Theory Workshop}, Oct. 2004.

\bibitem{WebAndJin06}
S.~Weber, J.~Andrews, and N.~Jindal, ``Ad hoc networks: the effect of fading,
  power control, and fully-distributed scheduling,'' submitted to \textit{IEEE
  Trans. Inform. Theory}, Dec. 2006.

\bibitem{Alouini_Goldsmith_ASE}
M.~S. Alouini and A.~Goldsmith, ``Area spectral efficiency of cellular mobile
  radio systems,'' \emph{IEEE Trans. Vehic. Tech.}, vol.~48, pp. 1047--1066,
  July 1999.

\end{thebibliography}

\end{document}